\documentclass[12pt,epsfig]{iopart}
\usepackage{epsfig}

\newlength{\dinwidth}
\newlength{\dinmargin}
\def\gsim{\,\lower.25ex\hbox{$\scriptstyle\sim$}\kern-1.30ex%
\raise 0.55ex\hbox{$\scriptstyle >$}\,}
\def\lsim{\,\lower.25ex\hbox{$\scriptstyle\sim$}\kern-1.30ex%
\raise 0.55ex\hbox{$\scriptstyle <$}\,}

\begin{document}
% Journal identifier can be put here if required, e.g.
%\jl{14}

\title{Outstanding problems in the phenomenology of hard diffractive scattering}

\author{B. E. Cox$^a$, K. Goulianos$^b$, L. L\"onnblad$^c$  and J. J. Whitmore$^d$}

\address{a) Department of Physics and Astronomy, University of Manchester, 
Manchester, M13 9PL, England. \\
b) The Rockefeller University, 
1230 York Avenue, New York,
NY 10021, USA. \\
c) Dept.~of Theoretical Physics 2, S\"olvegatan 14A, 
    S-223 62  Lund, Sweden \\
d) Department of Physics, 
Pennsylvania State University, 
104 Davey Laboratory, University Park,
Pennsylvania 16802-6300. \\}

\begin{abstract}
This paper is a summary of the discussion within the Diffractive and 
Low-x Physics Working Group at the 1999 Durham Collider Workshop of
 the interpretation of the Tevatron and HERA measurements of inclusive 
hard diffraction.
\end{abstract}
%\pacs{12.40.Nn, 13.60.Hb}
\section{The problems}

Although it has long been suspected that the factorisation of the hard 
component of diffractive scattering should not apply for hadron-hadron 
collisions, the magnitude of the breakdown at the Tevatron has come as a 
surprise. A model for diffractive hard scattering that contains all 
the essential features of the factorisation hypothesis is that of Ingelman 
and Schlein \cite{IS}. In diffractive DIS at HERA, for example, 
large rapidity gap events can be interpreted as the result of a highly 
virtual photon probing the structure of a pomeron `emitted' from 
the proton. Collins has recently proved factorisation for lepton induced 
diffractive processes \cite{collins}, but as expected the proof is not valid 
for hadron-hadron collisions. Alvero and collaborators \cite{ACW} 
have quantified this breakdown by extracting diffractive parton 
densities (which in the Ingelman-Schlein picture would be interpreted 
as parton distributions of the pomeron) from the HERA diffractive DIS 
and diffractive jet photoproduction data and using them to predict 
diffractive jet, W, Z and charm production and double pomeron exchange 
rates in $p \bar p $ collisions at the Tevatron. The diffractive parton 
distributions themselves are found to need a large amount of glue at the 
starting scale in order to fit the HERA photoproduction data. Hard gluon 
distributions ($1-\beta$ at large $\beta$) are preferred, although the 
present data cannot rule out an even harder distribution, similar to the 
form presented by the H1 Collaboration in their analysis of the diffractive 
DIS data \cite{H1diff}, that is strongly peaked towards $\beta = 1$. 
The predicted cross sections for the above processes at the Tevatron are 
consistently larger than those measured, the differences ranging from 
factors of a few for diffractive W and Z production to factors of up to 
30 for the gluon dominated fits in dijet production, indicating a severe 
breakdown of factorisation. The breakdown in the double pomeron rate is 
even more severe, where the gluon dominated fits fail by factors of order 100. For all processes, the low-glue fits, that are unfavoured at HERA, 
yield much better results. 
With this background, several questions present themselves. Firstly, 
is the picture of diffraction a la Ingelman-Schlein valid ? 
Are the parton distributions extracted at HERA any use outside HERA ? 
Should a new approach be sought ? Whilst not providing any answers, 
we will review several suggestions that may provide a starting point 
for future work.

\section{Possible Solutions}

\subsection{Rapidity Gap Survival Probability}

Perhaps the most obvious solution to the apparent low yield of rapidity gap 
events at the Tevatron relative to HERA is to attribute the difference to 
a rapidity gap survival factor. Such factors are able to explain the 
qualitative differences in rapidity gaps between jets fractions 
in $\sim$ 200 GeV $\gamma p$ collisions at HERA $(\sim 10 \%)$ and 
in 630 GeV $(\sim 3\%)$ and 1800 GeV $(\sim 1\%)$ $p \bar p$ collisions 
at the Tevatron \cite{CFL}, although large uncertainties remain. 
The idea is simple, although the creation of viable models is an extremely 
difficult problem \cite{CFL, GLM}. A rapidity gap produced by the 
exchange of a colour-singlet object may be filled in by secondary 
interactions between spectator partons in the event. Since there are more 
spectator partons in $p \bar p$ collisions than in $\gamma p$ collisions, 
and the number density of partons increases with increasing centre of mass 
energy, one would expect the rate of gap destruction to be significantly 
larger at the Tevatron than at HERA, and to increase with centre of mass 
energy. Such an analysis has yet to be performed for the case of hard 
diffractive scattering \cite{CFLP}. It is worth noting that, certainly in the gaps between jets case, 
it may be possible to control the non-perturbative physics by a careful 
definition of a rapidity gap. For example, a gap event may be defined as 
an event in which the total energy in a given rapidity region is greater 
than some value $Q$, where $Q \gg \Lambda_{QCD}$ \cite{OS}, or as an 
event in which there is no jet with $E_T^{jet} > E_{T(min)}$ in some 
rapidity region \cite{CFL}. 

An interesting question to ask in the context of gap survival is whether 
or not the gap destruction mechanism depends on the gap production 
subprocess. Most models to date introduce gap survival as a multiplicative 
factor dependent only on centre of mass energy, although this is almost 
certainly an over-simplification. In which case, the shapes of the 
diffractive parton distributions measured at different colliders and 
centre of mass energies would necessarily be different. Such a difference 
is present in the $\beta$ distributions measured by CDF 
and those extracted by H1 in diffractive DIS.

CDF measured the diffractive structure function of the antiproton 
using a method employing two samples of dijet events produced in $p \bar p$ 
collisions at $\sqrt s=1800$ GeV: a diffractive 
sample, collected  by triggering on a leading antiproton detected in 
a forward Roman Pot Spectrometer (RPS),
and an inclusive sample, collected with a minimum bias trigger.
In leading order QCD, the  ratio of the diffractive to inclusive 
cross sections as a function of the Bjorken $x$ of the struck parton 
of the antiproton, obtained 
from the dijet kinematics, is equal to the ratio of the 
corresponding structure functions. Thus, the diffractive structure can be 
calculated by multiplying the measured 
ratio of cross sections by the known inclusive 
structure. This method, which  bypasses the use of (model dependent) 
Monte Carlo generators, yields the colour-weighted structure function 
$$F_{jj}^{D}(x)=x\left\{g^D(x)
+\frac{4}{9}\sum_i\left[q_i^D(x)+{\bar q}_i^D(x)\right]\right\}$$
\noindent where $g^D(x)$ and $q^D(x)$ are the antiproton 
gluon and quark diffractive parton densities. Changing variables 
from $x$ to $\beta=x/\xi$, where $\xi$ is the $\bar p$ fractional momentum loss 
measured by the RPS, yields the structure function $F_{jj}^{D(3)}(\xi,\beta,Q^2)$.
\begin{figure}[h]
\centerline{\epsfig{figure=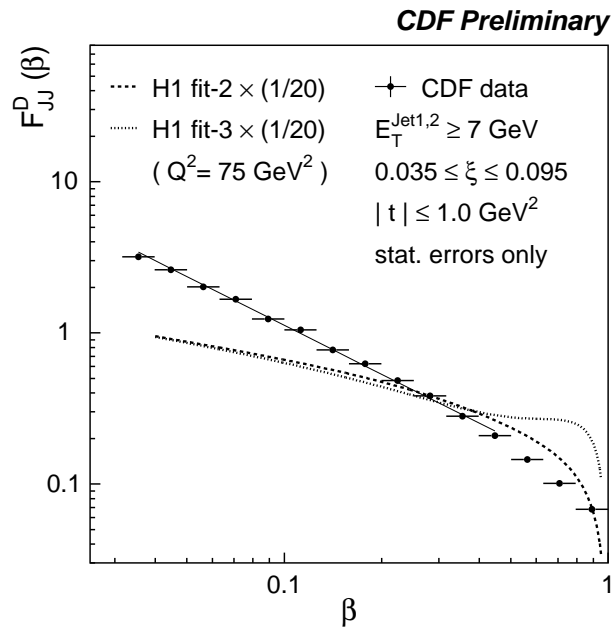,width=3.5in}}
\caption{CDF $\beta$ distribution (points) and fit (solid line) 
compared with expectations based on diffractive parton densities extracted 
from diffractive DIS measurements by the H1 Collaboration 
(dashed: fit2, dotted: fit 3).}
\label{fig:CDF}
\end{figure}
In Fig.~\ref{fig:CDF}, the $F_{jj}^{D(3)}(\xi,\beta,Q^2)$
measured by CDF at the Tevatron for 
$\langle Q^2\rangle\approx 75$ GeV$^2$ in the region of 
$0.035<\xi<0.095$ and  $|t|<1$ GeV$^2$ 
is compared (see~\cite{re:CDF}) with that calculated using parton densities 
extracted by the H1 Collaboration 
from  diffractive DIS measurements at HERA, scaled down by a factor of 20. 
The Tevatron and HERA $\beta$ distributions 
disagree both in normalisation and shape.
One should note, however, that the H1 data are in a different $\xi$ region to the Tevatron data, namely $\xi < 0.04$.  
In the H1 analysis, in which the data are fitted with two components, 
pomeron and reggeon, there are significant reggeon contributions in 
the $\xi$ region of the Tevatron data. For the reggeon, a pion structure 
function is assumed by H1. Allowing for a different reggeon structure, 
constrained by the data, could introduce some flexibility in the gluon 
component extracted from the diffractive DIS measurements.
Furthermore, one should note that the Tevatron data are mostly sensitive to
 the diffractive gluon content which, in the
 diffractive parton densities published by the H1
 collaboration\cite{H1diff},
 is derived from  the observed scaling violations of the diffractive
 structure function.
The question awaiting an answer must then be, is there a common set of 
diffractive pdf's which will fit in shape {\it both} the HERA and Tevatron measurements, 
leaving an overall normalisation factor which can be explained by a simple 
factorisable gap survival factor. Even if this is not so, can a sufficiently 
refined gap survival model account for the differences in the shape of the 
extracted diffractive pdf's? Or is a more fundamental revision of 
diffractive phenomenology called for?

\subsection{Soft Colour Interactions} 

The soft colour interaction approach \cite{SCI,GAL} differs from the 
above phenomenology in that it moves the gap formation from the 
initial state to the hadronisation phase. The hard subprocess and the 
perturbative evolution of partons is treated exactly the same for gap 
and non-gap events. However, after the perturbative phase, the colour 
structure of an event can be rearranged by exchange of soft gluons, 
typically between the perturbatively produced partons and the 
\textit{background colour field} of the hadrons. Such colour 
reconnections may result in regions devoid of colour, i.e.\ rapidity 
gaps. A review can be found in these proceedings \cite{LR}, where 
it is shown that fixing some global parameter for the reconnection 
probability to describe DIS gap events at HERA, it is not only 
possible to describe diffractive jet production (single \textbf{and} 
double diffraction) and diffractive W production at the Tevatron, but 
also a good description of high-$p_\perp$ quarkonium production is 
obtained. 

Some questionable features of the soft colour interaction models were 
pointed out during the workshop. The fact that the models do not 
modify in anyway the perturbative evolution of an event, means that 
e.g.\ the size of a rapidity gap in diffractive DIS events is 
completely determined by the most forward parton emitted in the 
perturbative phase. But so far the reproduction of DIS gap events has 
only been possible when implementing the soft colour interaction in 
the L\scalebox{0.8}{EPTO} generator \cite{LEPTO} which is known not to 
be able to describe perturbative emissions in the forward region (see 
e.g.\ \cite{H1forward}). Also, it has been shown \cite{LLGap} that 
introducing a similar reconnection model in the 
A\scalebox{0.8}{RIADNE} \cite{ariadne} program -- which \emph{is} able 
to describe perturbative features of the forward region -- neither the 
rate nor the distribution of rapidity gap events can be adequately 
described. The rate could, of course, be fixed by modifying the 
cut-off in the perturbative cascade or the reconnection probability, 
but this would not change the fact that e.g.\ the $m_X$ distribution 
comes out completely wrong. 

Although this casts serious doubts on the physical relevance of the 
soft colour interaction models, it does not prove that they are wrong. 
But to prove that they have anything to do with physics, it is highly 
desirable that they be implemented in an event generator which gives 
a reasonable description of the perturbative emissions in the forward 
region. After the workshop, work has started \cite{RikardWork} to 
implement soft colour interactions in the 
R\scalebox{0.8}{AP}G\scalebox{0.8}{AP} \cite{rapgap} program, which is 
similar to L\scalebox{0.8}{EPTO} but implements a \textit{resolved 
virtual photon} model to obtain a good description of e.g.\ forward 
jet rates. The model must also describe all diffractive HERA data. (i.e. charm, jets, photoproduction, etc). 

\subsection{A new approach}

A totally new approach has been suggested by one of us which avoids the 
above complications regarding gap survival, and allows the structure 
of the pomeron to be derived from that of the parent hadron.  Using a non-factorizing ansatz for $F_2^{D(3)}$, inspired by the 
observed scaling behavior of the soft single diffractive cross 
section\cite{Dino} 
\begin{equation} 
\label{eq:sd} 
\frac{d\sigma_{sd}}{dM^2}\propto \frac{1}{(M^2)^{1+\epsilon}} 
\end{equation} 
a formula is obtained which can be interpreted as a renormalized 
pomeron flux folded with the structure function $F_2$ of the proton in the 
pomeron--proton scattering subsystem. 
Details can be found in these proceedings \cite{dino}.

\section{Outlook} 

         Hard diffraction at HERA and the Tevatron is clearly not fully 
understood. The factorized pomerom picture does not explain all the data,
and whether or not any of the alternative models suggested will 
be fully successful remains to be seen. This difficult border region between 
perturbative and non-perturbative QCD remains a challenge, which will 
probably require more experimental data before it can be met successfully.

\end{document}